\documentclass[conference]{IEEEtran}
\IEEEoverridecommandlockouts
\usepackage[utf8]{inputenc}
\usepackage[T1]{fontenc}
\usepackage{amsmath,amsfonts}

\ifCLASSINFOpdf
\else
   \usepackage[dvips]{graphicx}
\fi

\usepackage{cite}
\usepackage{svg}

\usepackage[ruled,vlined]{algorithm2e}
\usepackage{algorithmic}
\usepackage{listings,lstautogobble}
\usepackage{mathtools}

\usepackage{ragged2e}
\usepackage{array}
\usepackage{ltxtable}
\usepackage{booktabs}
\usepackage{supertabular}

\usepackage{array}
\usepackage{textcomp}
\usepackage{url}
\usepackage{verbatim}
\usepackage{graphicx}
\usepackage{cite}
\usepackage{xcolor}
\begin{document}

\title{Joint Downlink-Uplink Channel Estimation for Non-Reciprocal RIS-Assisted Communications
\thanks{The authors acknowledge the partial support of Fundação Cearense de Apoio ao Desenvolvimento Científico e Tecnológico (FUNCAP) under grants FC3-00198-00056.01.00/22 and  ITR-0214-00041.01.00/23, the National Institute of Science and Technology (INCT-Signals) sponsored by Brazil's National Council for Scientific and Technological Development (CNPq) under grant 406517/2022-3, and the Coordenação de Aperfeiçoamento de Pessoal de Nível Superior - Brasil (CAPES). This work is also partially supported by CNPq under grants 312491/2020-4 and 443272/2023-9.}
}

\author{\IEEEauthorblockN{Paulo R. B. Gomes\IEEEauthorrefmark{2}\IEEEauthorrefmark{1}, Amarilton L. Magalhães\IEEEauthorrefmark{2}\IEEEauthorrefmark{1}, André L. F. de Almeida\IEEEauthorrefmark{2}}\\
\IEEEauthorblockA{Federal University of Ceará\IEEEauthorrefmark{2}, Federal Institute of Education, Science and Technology of Ceará\IEEEauthorrefmark{1}, Fortaleza, Brazil
\\E-mail: gomes.paulo@ifce.edu.br, amarilton@gtel.ufc.br, andre@gtel.ufc.br}
}

\maketitle

\begin{abstract}
Reconfigurable intelligent surface (RIS) is a recent low-cost and energy-efficient technology with potential applicability for future wireless communications. Performance gains achieved by employing RIS directly depend on accurate channel estimation (CE). It is common in the literature to assume channel reciprocity due to the facilities provided by this assumption, such as no channel feedback, beamforming simplification, and latency reduction. However, in practice, due to hardware limitations at the RIS and transceivers, the channel non-reciprocity may occur naturally, so such behavior needs to be considered. In this paper, we focus on the CE problem in a non-reciprocal RIS-assisted multiple-input multiple-output (MIMO) wireless communication system. Making use of a novel closed-loop three-phase protocol for non-reciprocal CE estimation, we propose a two-stage fourth-order
Tucker decomposition-based CE algorithm. In contrast to classical time-division duplexing (TDD) and frequency-division duplexing (FDD) approaches the proposed method concentrates all the processing burden for CE on the base station (BS) side, thereby freeing hardware-limited user terminal (UT) from this task. Our simulation results show that the proposed method has satisfactory performance in terms of CE accuracy compared to benchmark FDD LS-based and tensor-based techniques.
\end{abstract}

\begin{IEEEkeywords}
RIS, MIMO, non-reciprocity, channel estimation, Tucker decomposition, TALS, KRF.
\end{IEEEkeywords}

\section{Introduction}
Reconfigurable intelligent surface (RIS) technology has been receiving wide notoriety since its employment provides transformative advances for the next generation of wireless communications. RISs are manufactured software-controlled structures composed of a large number of hardware-efficient (passive/semi-passive) reconfigurable scattering elements to reflect the incident electromagnetic waves (EM) in a proper manner when configured intelligently \cite{8796365}. By reflecting the EM, the uncontrollable propagation environment is then replaced by a ``\textit{smart environment}'' assisted by the RIS. Their ability to adaptively optimize the EM propagation presents advantages, allowing for energy efficiency, low latency, higher data rates, massive connectivity, and extended coverage without requiring substantial infrastructure upgrades \cite{9200578}. The benefits of RISs are fully established when accurate channel estimates are achieved since channel knowledge allows the RIS controller to properly design the passive RIS phase-shifts following some optimization criterion \cite{10403475, 10690066}. Several works have provided solutions to solve the channel estimation (CE) problem in RIS-assisted communication systems \cite{9722893}. 

Tensor decompositions have proven over the years to be powerful tools with different applications in the signal processing field \cite{7891546}, including wireless communications \cite{DEALMEIDA2007337, 4524041, 6897995, 7152972}. In the context of multiple-input multiple-output (MIMO) RIS-assisted wireless communication systems, the authors of \cite{9361077} proposed two iterative and closed-form pilot-assisted CE methods based on the well-known \textit{parallel factor} (PARAFAC) tensor decomposition. The tensor-based methods proposed by \cite{9361077} outperformed traditional LS-based solutions in terms of CE accuracy. In \cite{10137372}, CE is addressed in a scenario in which the RIS is subject to physical/hardware limitations and environmental impairments. 

Most of the existing works assume channel reciprocity. However, in reality, due to hardware limitations at the RIS and transceivers, the non-reciprocity of the involved channels may occur naturally, so such behavior needs to be considered in the modeling. This makes the CE problem more challenging since it involves the estimation of multiple channels simultaneously: two separate downlink (DL) and uplink (UL) channels totaling four channels to be estimated. A scarce number of works have focused on the CE problem for non-reciprocal RIS-assisted wireless communications. The authors of \cite{10201815} proposed an alternating algorithm for CE in a non-reciprocal single-input single-output (SISO) scenario. 

In this work, we investigate CE in non-reciprocal MIMO RIS-assisted wireless communication systems. A novel three-phase closed-loop CE protocol is formulated. This novel protocol allows to concentrate the processing burden for CE at the base-station (BS) side and relaxes both channel and RIS non-reciprocities \cite{9690479, 9099621} while avoiding complex signal processing tasks for CE at hardware-limited user-terminal (UT) side. Based on the protocol, we develop a fourth-order Tucker decomposition-based approach for joint DL and UL CE. The proposed method solves the CE of non-reciprocal channels as a unified problem in two stages. In the first one, estimates of non-reciprocal (DL and UL) channels between BS-RIS are obtained iteratively by means of the well-known trilinear alternating least square (TALS) procedure. In the second one, the RIS-UT DL and UL channels are jointly estimated through a closed-form way from the Khatri-Rao factorization (KRF) method. Simulation results validate the accuracy of the proposed method compared to frequency-division duplexing (FDD) LS-based and tensor-based techniques.

\subsection{Notation and properties}
Scalars, column vectors, matrices, and tensors are represented by $a$, $\mathbf{a}$, $\mathbf{A}$ and $\mathcal{A}$, respectively. The superscripts $(\cdot)^{\text{T}}$, $(\cdot)^{*}$ and $(\cdot)^{\dag}$ represent the transpose, conjugate, and Moore-Penrose pseudo-inverse, respectively. $\|\cdot\|_{\text{F}}$ is the Frobenius norm. An $K \times K$ identity matrix is denoted by $\mathbf{I}_{K}$, while an $N$-way identity tensor of size $R \times R \cdots \times R$ is $\mathcal{I}_{N,R}$. $\text{diag}(\mathbf{a})$ converts $\mathbf{a} \in \mathbb{C}^{I \times 1}$ to a $I \times I$ diagonal matrix, while $D_{i}(\mathbf{A})$ forms a diagonal matrix of size $R \times R$ from the $i$-th row of $\mathbf{A} \in \mathbb{C}^{I \times R}$. $\text{vec}(\mathbf{A})$ vectorizes $\mathbf{A}$ to $\mathbf{a} \in \mathbb{C}^{IR \times 1}$. In the opposite way, $\text{unvec}_{I \times R}(\mathbf{a})$ returns the vector argument to $\mathbf{A} \in \mathbb{C}^{I \times R}$. The symbols $\circ$, $\otimes$, and $\diamond$ denote the outer product, Kronecker and Khatri-Rao products, respectively. In this paper, we shall make use of the following identities:
\begin{align}
\label{kk}\mathbf{A}\mathbf{C} \diamond \mathbf{B}\mathbf{D} &= (\mathbf{A} \otimes \mathbf{B})(\mathbf{C} \diamond \mathbf{D}),\\
\label{p1}\text{vec}(\mathbf{A}\mathbf{B}\mathbf{C}) &= (\mathbf{C}^{\text{T}} \otimes \mathbf{A})\text{vec}(\mathbf{B}),\\
\label{p3} \text{vec}(\mathbf{A}\mathbf{B}\mathbf{C}) &= (\mathbf{C}^{\text{T}} \diamond \mathbf{A})\text{vecd}(\mathbf{B}), ~(\mathbf{B} \mathrm{\mbox{ diagonal}})\\
\label{p2}\text{diag}(\mathbf{a})\mathbf{b} &= \text{diag}(\mathbf{b})\mathbf{a},
\end{align}
where the involved vectors and matrices have compatible dimensions in each case. Moreover, the definitions and useful operations involving tensors are in accordance with \cite{kolda2009tensor}. 

\begin{figure}[t]
\centering
\includegraphics[width=0.45\textwidth]{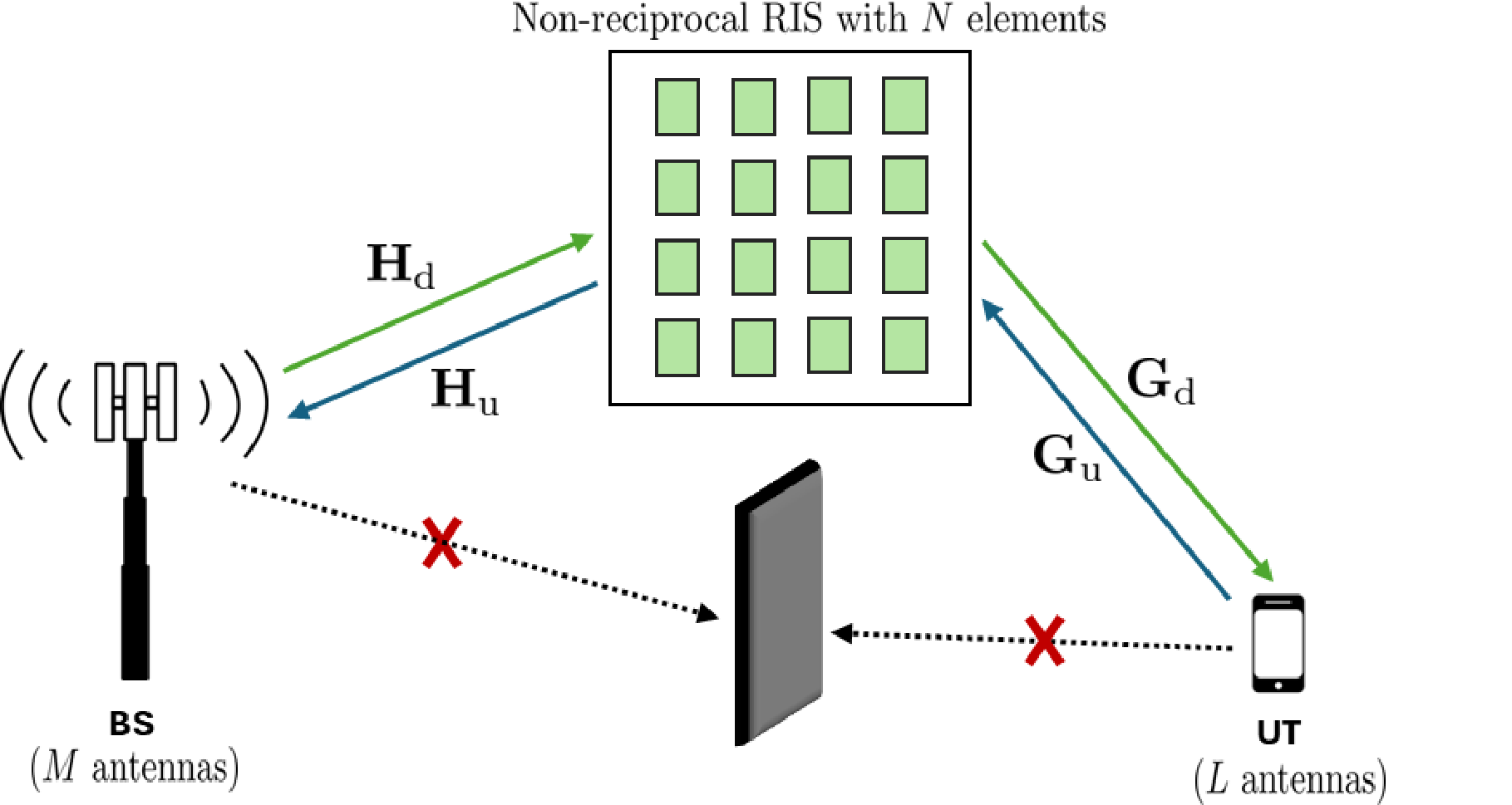}
\caption{\small{Non-reciprocal RIS-assisted MIMO wireless communication system. 
Non-reciprocity of the channels is considered for all MIMO channels involved in the communication.}}
\label{scenario}
\end{figure}

\section{System and Signal Models}\label{systemModel}
\subsection{System Description}\label{systemDescription}
In Fig. \ref{scenario}, we consider a MIMO wireless communication system assisted by a passive RIS composed of $N$ individually adjustable scattering elements. 
The BS and UT are equipped with $M$ and $L$ antennas, respectively. We assume that the direct link between the BS and UT (in DL and UL communications) are blocked due to obstacles or sufficiently weak due to deep fading, thus they are not considered in our signal models. Following a more general and realistic representation, we assume a non-reciprocal system, i.e., a non-reciprocal RIS, is deployed, and non-reciprocal channels with different fading coefficients are considered. In the DL, $\mathbf{H}_{\text{d}} \in \mathbb{C}^{M \times N}$ and $\mathbf{G}_{\text{d}} \in \mathbb{C}^{L \times N}$ denote the MIMO channel matrices from the BS-RIS and RIS-UT, respectively. Similarly, $\mathbf{H}_{\text{u}} \in \mathbb{C}^{M \times N}$ and $\mathbf{G}_{\text{u}} \in \mathbb{C}^{L \times N}$ denote the MIMO channel matrices from the RIS-BS and UT-RIS in the UL, respectively. Note that, due to the non-reciprocity of the channels $\mathbf{H}_{\text{d}} \neq \mathbf{H}_{\text{u}}$ and $\mathbf{G}_{\text{d}} \neq \mathbf{G}_{\text{u}}$\footnote{Although the literature conventionally uses matrix transposition to indicate channel reciprocity, this representation is omitted here in order to simplify the notation in our tensor modeling which will be detailed in Section \ref{tensorModeling}.}. Therefore, the symmetry between the DL and UL channels, a fundamental property in time-division duplexing (TDD) schemes, does not hold, invalidating the joint active (BS and UT) and passive (RIS) beamforming designs in the UL reusing the DL CSI (or in the DL reusing the UL CSI).

\subsection{Proposed Closed-Loop CE Protocol}\label{protocolDescription}
Our proposed CE protocol assumes that DL and UL channels in Fig. \ref{scenario} are flat quasi-static block fading and remain constant during the coherence time, varying independently from different coherence intervals. The CE protocol is divided into three phases detailed below:

\textit{Phase 1}: A two-timescale pilot transmission protocol is adopted. It consists of $K$ blocks, each of them having a duration of $T$ time slots, totaling a time interval of $KT$ symbol periods \cite{9361077}. At each individual block $k = 1, \ldots, K$ the same length-$T$ pilot sequence is repeatedly transmitted while the DL RIS response $\mathbf{s}_{\text{d}}[k] \in \mathbb{C}^{N \times 1}$ remains constant within the $k$-th block but varies among blocks, yielding a total of $\mathbf{s}_{\text{d}}[1], \ldots, \mathbf{s}_{\text{d}}[K]$ different RIS responses. The DL RIS response $\mathbf{s}_{\text{d}}[k] \in \mathbb{C}^{N \times 1}$ at the $k$-th block is defined as $\mathbf{s}_{\text{d}}[k] = [\alpha_{1,k}e^{j\phi_{1,k}}, \ldots, \alpha_{N,k}e^{j\phi_{N,k}}]^{\text{T}} \in \mathbb{C}^{N \times 1}$, where $\phi_{n,k} \in (0, 2\pi]$ and $\alpha_{n,k} \in \{0,1\}$ denote the phase shift and the amplitude scattering coefficient of the $n$-th RIS element adjusted at the $k$-th block, respectively. 

\textit{Phase 2}: Motivated by the non-reciprocity of the system and in contrast to typical TDD and FDD schemes, the UT (that naturally may be a hardware-limited device) does not perform CE processing. Instead, a simple coding strategy is applied by the UT, reducing its hardware complexity. For each received block $k = 1, \ldots, K$, the UT will apply $p = 1, \ldots, P$ different linear coding per UT antenna using the Khatri-Rao coding proposed by \cite{1033671}. Some level of coordination is required between the BS and UT since coding must be known at the BS to perform joint DL and UL CE. 

\textit{Phase 3}: At the UT the $p = 1, \ldots, P$ coded pilot signals are fed back to the BS. In the UL, the same procedure used for \textit{Phase 1} is carried out, with the difference that the RIS switches over $\mathbf{s}_{\text{u}}[1], \ldots, \mathbf{s}_{\text{u}}[K]$ scattering patterns, Different RIS responses in the DL and UL model the RIS non-reciprocity \cite{9690479,9099621}. Note that $\mathbf{s}_{\text{u}}[k] \in \mathbb{C}^{N \times 1}$ denotes the UL RIS response at the $k$-th block, which has a similar structure to $\mathbf{s}_{\text{d}}[k]$ but with different amplitude and phase parameters due to non-reciprocity. After this closed-loop transmission, the BS (which has greater processing capability compared to UT) can jointly estimate the non-reciprocal channels $\{\mathbf{H}_{\text{d}}$, $\mathbf{G}_{\text{d}}$, $\mathbf{H}_{\text{u}}$, $\mathbf{G}_{\text{u}}\}$ without overloading the UT. 

\subsection{Closed-Loop Signal Model}
Following the proposed CE protocol, the BS transmits the length-$T$ pilot sequence matrix $\mathbf{X} = [\mathbf{x}[1], \ldots, \mathbf{x}[T]] \in \mathbb{C}^{M \times T}$ per block $k = 1,\ldots,K$, where $\mathbf{x}[t] \in \mathbb{C}^{M \times 1}$ is the pilot sequence vector at the $t$-th time slot, $t=1,\ldots,T$. Thus, the discrete-time baseband signal received by the UT in the $t$-th time slot at the $k$-th block is given by
\begin{equation}\label{dlSignal}
\bar{\mathbf{y}}[t,k] = \mathbf{G}_{\text{d}}\text{diag}(\mathbf{s}_{\text{d}}[k])\mathbf{H}_{\text{d}}^{\text{T}}\mathbf{x}[t] + \mathbf{v}_{\text{d}}[t,k] \in \mathbb{C}^{L \times 1},
\end{equation}
where $\mathbf{v}_{\text{d}}[t,k] \in \mathbb{C}^{L \times 1}$ represents the additive white Gaussian noise (AWGN) at the UT. By collecting all the $t = 1, \ldots, T$ time slots within the $k$-th block, we recast the DL received signal (\ref{dlSignal}) in a matrix form as follows
\begin{equation}\label{dlMatrix}
\bar{\mathbf{\mathbf{Y}}}[k] = \mathbf{G}_{\text{d}}D_{k}(\mathbf{S}_{\text{d}})\mathbf{H}_{\text{d}}^{\text{T}}\mathbf{X} + \mathbf{V}_{\text{d}}[k] \in \mathbb{C}^{L \times T},
\end{equation}
where $\bar{\mathbf{Y}}[k] = [\bar{\mathbf{y}}[1,k], \ldots, \bar{\mathbf{y}}[T,k]] \in \mathbb{C}^{L \times T}$ and $\mathbf{V}_{\text{d}} = [\mathbf{v}_{\text{d}}[1,k], \ldots, \mathbf{v}_{\text{d}}[T,k]] \in \mathbb{C}^{L \times T}$. The matrix $\mathbf{S}_{\text{d}} =[\mathbf{s}_{\text{d}}[1], \ldots, \mathbf{s}_{\text{d}}[K]]^{\text{T}} \in \mathbb{C}^{K \times N}$ collects in its rows the $k = 1, \ldots, K$ DL RIS responses configured in \textit{Phase 1} of the CE protocol.

At the UT side, the linear coding performed in \textit{Phase 2} changes the received signal (\ref{dlMatrix}) in a coordinated manner. In other words, the $k$-th received block is repeated $P$ times, coded in each of them by means of a diagonal matrix $D_{p}(\mathbf{C}) \in \mathbb{C}^{L \times L}$, and then transmitted back to the BS. On the $p$-th time slot, the UT sends the following coded signal back to the BS
\begin{equation}\label{amplifiedSigna}
\bar{\mathbf{Y}}[k,p] = D_{p}(\mathbf{C})\bar{\mathbf{Y}}[k] \in \mathbb{C}^{L \times T},
\end{equation}
where $\mathbf{C} = [\mathbf{c}[1], \ldots, \mathbf{c}[P]]^{\text{T}} \in \mathbb{C}^{P \times L}$ is the known code matrix whose its $p$-th row contains the coding factors per UT antenna, $l = 1, \ldots, L$.

Finally, in \textit{Phase 3}, the coded signal (\ref{amplifiedSigna}) is reflected by the RIS towards the BS by means of $k = 1, \ldots, K$ scattering patterns. The UL received closed-loop signal at the BS is represented by 
\begin{equation}\label{bsSignal}
\bar{\mathbf{Q}}[k,p] = \mathbf{H}_{\text{u}}D_{k}(\mathbf{S}_{\text{u}})\mathbf{G}_{\text{u}}^{\text{T}}\bar{\mathbf{Y}}[k,p] + \mathbf{V}_{\text{u}}[k,p] \in \mathbb{C}^{M \times T},
\end{equation}
where $\mathbf{S}_{\text{u}} = [\mathbf{s}_{\text{u}}[1], \ldots, \mathbf{s}_{\text{u}}[K]]^{\text{T}} \in \mathbb{C}^{K \times N}$ is the UL RIS response, and $\mathbf{V}_{\text{u}}[k,p] \in \mathbb{C}^{M \times T}$ is the noise term at the BS.

\section{Tensor Signal Modeling}\label{tensorModeling}
In order to simplify our tensor formulation, we assume that the BS sends orthogonal pilot sequences, such that $\mathbf{X}\mathbf{X}^{\text{H}} = \mathbf{I}_{M}$. After a bilinear matched-filtering in (\ref{bsSignal}) using $\mathbf{X}^{\text{H}}$, the filtered version of the received signal, denoted by $\mathbf{Q}[k,p] = \bar{\mathbf{Q}}[k,p]\mathbf{X}^{\text{H}} \in \mathbb{C}^{M \times M}$, can be written as
\begin{equation}\label{signalPart}
\mathbf{Q}[k,p] = \underbrace{\mathbf{H}_{\text{u}}D_{k}(\mathbf{S}_{\text{u}})\mathbf{G}_{\text{u}}^{\text{T}}D_{p}(\mathbf{C})\mathbf{G}_{\text{d}}D_{k}(\mathbf{S}_{\text{d}})\mathbf{H}_{\text{d}}^{\text{T}}}_{\mathbf{Z}[k,p] \in \mathbb{C}^{M \times M}} + \mathbf{V}[k,p],
\end{equation}
where the first term on the right side $\mathbf{Z}[k,p] \in \mathbb{C}^{M \times M}$ represents the useful signal part for our model (i.e., a unified representation of the DL and UL signals coming from the closed-loop CE protocol), while the second term $\mathbf{V}[k,p] = \mathbf{H}_{\text{u}}D_{k}(\mathbf{S}_{\text{u}})\mathbf{G}_{\text{u}}^{\text{T}}D_{p}(\mathbf{C})\mathbf{V}_{\text{d}}[k]\mathbf{X}^{\text{H}} + \mathbf{V}_{\text{u}}[k,p]\mathbf{X}^{\text{H}} \in \mathbb{C}^{M \times M}$ denotes the overall noise contribution.

By processing (\ref{signalPart}) when BS collects the UL signal across all $p = 1, \ldots, P$ retransmissions for all $k = 1, \ldots, K$ blocks a fourth-order tensor $\mathcal{Q} = \mathcal{Z} + \mathcal{V} \in \mathbb{C}^{M \times M \times \times K \times L}$ is obtained. For this, the $\text{vec}(\cdot)$ operator is applied in (\ref{signalPart}) together with properties (\ref{p1}) and (\ref{p2}), resulting in $\mathbf{q}[k,p] = \mathbf{z}[k,p] + \mathbf{v}[k,p] \in \mathbb{C}^{M^{2} \times 1}$ denoted by
\begin{align*}
\mathbf{q}[k,p] \notag&= \left(\mathbf{H}_{\text{d}} \otimes \mathbf{H}_{\text{u}}\right)\text{vec}\left(D_{k}(\mathbf{S}_{\text{u}})\mathbf{W}_{p}D_{k}(\mathbf{S}_{\text{d}})\right) + \mathbf{v}[k,p] \\
\notag&= \left(\mathbf{H}_{\text{d}} \otimes \mathbf{H}_{\text{u}}\right)\left(D_{k}(\mathbf{S}_{\text{d}}) \otimes D_{k}(\mathbf{S}_{\text{u}})\right)\mathbf{w}_{p} + \mathbf{v}[k,p] \\
&= \left(\mathbf{H}_{\text{d}} \otimes \mathbf{H}_{\text{u}}\right)\text{diag}(\mathbf{w}_{p})\mathbf{S}(k,:)^{\text{T}} + \mathbf{v}[k,p],
\end{align*}
where 
$\mathbf{q}[k,p] = \text{vec}\left(\mathbf{Q}[k,p]\right) \in \mathbb{C}^{M^{2} \times 1}$, $\mathbf{z}[k,p] = \text{vec}\left(\mathbf{Z}[k,p]\right) \in \mathbb{C}^{M^{2} \times 1}$, $\mathbf{v}[k,p] = \text{vec}\left(\mathbf{V}[k,p]\right) \in \mathbb{C}^{M^{2} \times 1}$, $\mathbf{W}_{p} = \mathbf{G}_{\text{u}}^{\text{T}}D_{p}(\mathbf{C})\mathbf{G}_{\text{d}} \in \mathbb{C}^{N \times N}$, $\mathbf{w}_{p} = \text{vec}(\mathbf{W}_{p}) \in \mathbb{C}^{N^{2} \times 1}$, and $\mathbf{S}(k,:) = \mathbf{S}_{\text{d}}(k,:) \otimes \mathbf{S}_{\text{u}}(k,:) \in \mathbb{C}^{1 \times N^{2}}$ is the combined DL/UL RIS response associated with $k$-th block.

Collecting $\mathbf{q}[k,p]$ for all the $k = 1, \ldots, K$ blocks that form the $p$-th retransmission slot as the columns of the resulting matrix $\mathbf{Q}[p] = \left[\mathbf{q}[1,p], \ldots, \mathbf{q}[K,p]\right] \in \mathbb{C}^{M^{2} \times K}$, we obtain
\begin{align}\label{vecP}
\mathbf{Q}[p] \notag&= \left(\mathbf{H}_{\text{d}} \otimes \mathbf{H}_{\text{u}}\right)\text{diag}(\mathbf{w}_{p})\left[\mathbf{S}(1,:)^{\text{T}}, \ldots, \mathbf{S}(k,:)^{\text{T}}\right] \\
\notag&\quad\quad+ \left[\mathbf{v}[1,p], \ldots, \mathbf{v}[K,p]\right] \\ &= \left(\mathbf{H}_{\text{d}} \otimes \mathbf{H}_{\text{u}}\right)\text{diag}(\mathbf{w}_{p})\mathbf{S}^{\text{T}} + \mathbf{V}[p],
\end{align}
where $\mathbf{S} = (\mathbf{S}_{\text{d}}^{\text{T}} \diamond \mathbf{S}_{\text{u}}^{\text{T}})^{\text{T}} \in \mathbb{C}^{K \times N^{2}}$ unifies the DL and UL RIS responses in the closed-loop transmission across all the $K$ blocks as a Khatri-Rao structure, and $\mathbf{V}[p] = \left[\mathbf{v}[1,p], \ldots, \mathbf{v}[K,p]\right] \in \mathbb{C}^{M^{2} \times K}$.

Now, applying the $\text{vec}(\cdot)$ operator and property (\ref{p3}) to (\ref{vecP}) we obtain $\mathbf{q}[p] = \left[\mathbf{S} \diamond \left(\mathbf{H}_{\text{d}} \otimes \mathbf{H}_{\text{u}}\right)\right]\mathbf{w}_{p} + \mathbf{v}[p]$, where $\mathbf{q}[p] = \text{vec}(\mathbf{Q}[p]) \in \mathbb{C}^{M^{2}K \times 1}$ and $\mathbf{v}[p] = \text{vec}(\mathbf{V}[p]) \in \mathbb{C}^{M^{2}K \times 1}$. Using $\mathbf{w}_{p} = \text{vec}(\mathbf{W}_{p})$, we can rewrite the vectorized signal $\mathbf{q}[p]$ in the following equivalent form
\begin{equation}
\mathbf{q}[p] \notag= \left[\mathbf{S} \diamond \left(\mathbf{H}_{\text{d}} \otimes \mathbf{H}_{\text{u}}\right)\right]\left(\mathbf{G}_{\text{d}}^{\text{T}} \diamond \mathbf{G}_{\text{u}}^{\text{T}}\right)\mathbf{C}(p,:)^{\text{T}} + \mathbf{v}[p].
\end{equation}
Finally, collecting all the $\mathbf{q}[p]$, $p = 1, \ldots, P$, as the columns of $\mathbf{Q} = [\mathbf{q}[1], \ldots, \mathbf{q}[P]] \in \mathbb{C}^{M^{2}K \times P}$ leads to
\begin{equation}\label{pmf}
\mathbf{Q} = \left[\mathbf{S} \diamond \left(\mathbf{H}_{\text{d}} \otimes \mathbf{H}_{\text{u}}\right)\right]\left(\mathbf{G}_{\text{d}}^{\text{T}} \diamond \mathbf{G}_{\text{u}}^{\text{T}}\right)\mathbf{C}^{\text{T}} + \mathbf{V},
\end{equation}
where $\mathbf{V} = [\mathbf{v}[1], \ldots, \mathbf{v}[P]] \in \mathbb{C}^{M^{2}K \times P}$. Performing a bilinear matched-filtering by multiplying both sides of (\ref{pmf}) by the pseudo-inverse of $\mathbf{C}^{\text{T}}$ (when $P \geq L$), results in
\begin{equation}\label{tot}
\tilde{\mathbf{Q}} = \left[\mathbf{S} \diamond \left(\mathbf{H}_{\text{d}} \otimes \mathbf{H}_{\text{u}}\right)\right]\left(\mathbf{G}_{\text{d}}^{\text{T}} \diamond \mathbf{G}_{\text{u}}^{\text{T}}\right) + \tilde{\mathbf{V}}, 
\end{equation}
where $\tilde{\mathbf{Q}} = \mathbf{Q}\left(\mathbf{C}^{\text{T}}\right)^{\dag} \mathbb{C}^{M^{2}K \times L}$ and $\tilde{\mathbf{V}} = \mathbf{V}\left(\mathbf{C}^{\text{T}}\right)^{\dag} \in \mathbb{C}^{M^{2}K \times L}$, respectively. 

Let us define $\mathbf{G} = \mathbf{G}_{\text{d}}^{\text{T}} \diamond \mathbf{G}_{\text{u}}^{\text{T}} \in \mathbb{C}^{N^2 \times L}$. Then, $\tilde{\mathbf{Q}}^{\text{T}} = \mathbf{G}^{\text{T}}\left[\mathbf{S} \diamond \left(\mathbf{H}_{\text{d}} \otimes \mathbf{H}_{\text{u}}\right)\right]^{\text{T}} + \tilde{\mathbf{V}}^{\text{T}} \in \mathbb{C}^{L \times M^{2}K}$. According to \cite{kolda2009tensor}, the transpose of (\ref{tot}) corresponds to mode-3 unfolding of the third-order tensor $\tilde{\mathcal{Q}} = \tilde{\mathcal{Z}} + \tilde{\mathcal{V}} \in \mathbb{C}^{M^2 \times K \times L}$, i.e., $\tilde{\mathbf{Q}}^{\text{T}} = [\tilde{\mathcal{Q}}]_{(3)} \in \mathbb{C}^{L \times M^{2}K}$, in which
\begin{equation}\label{pf3}
\tilde{\mathcal{Q}} = \underbrace{\mathcal{I}_{3,N^2} \times_{1} \left(\mathbf{H}_{\text{d}} \otimes \mathbf{H}_{\text{u}}\right) \times_{2} \mathbf{S} \times_{3} \mathbf{G}^{\text{T}}}_{\tilde{\mathcal{Z}} \in \mathbb{C}^{M^2 \times K \times L}} + \tilde{\mathcal{V}},
\end{equation}
\vspace{+0.2cm}
where $\tilde{\mathcal{Z}} \in \mathbb{C}^{M^2 \times K \times L}$ and $\tilde{\mathcal{V}} \in \mathbb{C}^{M^2 \times K \times L}$ denote the useful signal tensor and the noise tensor obtained by reshaping the matrices $\tilde{\mathbf{Q}}^{\text{T}}$ and $\tilde{\mathbf{V}}^{\text{T}}$ as third-order tensors, respectively. Looking at (\ref{pf3}), we can observe that the tensor $\tilde{\mathcal{Q}}$ has size $M^2$ (related to the number of antennas in the BS) in its first dimension. To explore more of the tensor gain, we can decouple this dimension of high size into two smaller dimensions of size $M$ each by formulating $\tilde{\mathcal{Q}}$ in its alternative fourth-order representation form $M \times M \times K \times L$. Considering the mode-1 unfolding of $\tilde{\mathcal{Z}}$, i.e., $[\tilde{\mathcal{Z}}]_{(1)} = \left(\mathbf{H}_{\text{d}} \otimes \mathbf{H}_{\text{u}}\right)\left[\mathbf{G}^{\text{T}} \diamond \mathbf{S}\right]^{\text{T}} \in \mathbb{C}^{M^2 \times KL}$ and applying the Khatri-Rao product property in (\ref{kk}) yields
\begin{equation}
[\mathcal{Z}]_{([1,2],[3,4])} = (\mathbf{H}_{\text{d}} \otimes \mathbf{H}_{\text{u}})[\mathcal{R}]_{([1,2],[3,4])}\left(\mathbf{G}^{\text{T}} \otimes \mathbf{S}\right)^{\text{T}},
\end{equation}
where $[\mathcal{R}]_{([1,2],[3,4])} = \left(\mathbf{I}_{N^2} \diamond \mathbf{I}_{N^2}\right)^{\text{T}} \in \mathbb{C}^{N^2 \times N^4}$ denotes the multimode unfolding of the fourth-order tensor $\mathcal{R} \in \mathbb{C}^{N \times N \times N^2 \times N^2}$ that merges its first and second dimensions row-wise and the third and fourth dimensions column-wise \cite{8313190}. Similarly, $[\mathcal{Z}]_{([1,2],[3,4])} \in \mathbb{C}^{M^2 \times KL}$ represents the multimode unfolding of the fourth-order signal tensor $\mathcal{Z} \in \mathbb{C}^{M \times M \times K \times L}$ expressed in terms of mode-$n$ product as
\begin{equation}
\mathcal{Z} = \mathcal{R} \times_{1} \mathbf{H}_{\text{u}} \times_{2} \mathbf{H}_{\text{d}} \times_{3} \mathbf{S} \times_{4} \mathbf{G}^{\text{T}}.
\end{equation}
It is now clear that the noisy signal tensor in (\ref{pf3}) corresponds to the multimode representation of the fourth-order tensor $\mathcal{Q} = \mathcal{Z} + \mathcal{V} \in \mathbb{C}^{M \times M \times K \times L}$ where $\mathcal{Q}$, $\mathcal{Z}$ and $\mathcal{V}$ are obtained by decoupling the first dimension of $\tilde{\mathcal{Q}}$, $\tilde{\mathcal{Z}}$ and $\tilde{\mathcal{V}}$ into two independent dimensions of size $M$ each, i.e., converting them to fourth-order tensors. Therefore, the closed-loop received signal at the BS is finally obtained as
\begin{equation}\label{t4}
\mathcal{Q} = \mathcal{R} \times_{1} \mathbf{H}_{\text{u}} \times_{2} \mathbf{H}_{\text{d}} \times_{3} \mathbf{S} \times_{4} \mathbf{G}^{\text{T}} + \mathcal{V}.
\end{equation}

Equation (\ref{t4}) corresponds to the Tucker-4 tensor decomposition of $\mathcal{Q} \in \mathbb{C}^{M \times M \times K \times L}$. According to \cite{kolda2009tensor}, the four unfoldings of $\mathcal{Q}$, represented by $[\mathcal{Q}]_{(1)} \in \mathbb{C}^{M \times MKL}$, $[\mathcal{Q}]_{(2)} \in \mathbb{C}^{M \times MKL}$, $[\mathcal{Q}]_{(3)} \in \mathbb{C}^{K \times M^{2}L}$ and $[\mathcal{Q}]_{(4)} \in \mathbb{C}^{L \times M^{2}L}$, respectively, admit the following factorizations in terms of its factor matrices and core tensor
\begin{align}
\label{unf1}[\mathcal{Q}]_{(1)} &= \mathbf{H}_{\text{u}}[\mathcal{R}]_{(1)}\left(\mathbf{G}^{\text{T}} \otimes \mathbf{S} \otimes \mathbf{H}_{\text{d}}\right)^{\text{T}} + [\mathcal{V}]_{(1)}, \\
\label{unf2}[\mathcal{Q}]_{(2)} &= \mathbf{H}_{\text{d}}[\mathcal{R}]_{(2)}\left(\mathbf{G}^{\text{T}} \otimes \mathbf{S} \otimes \mathbf{H}_{\text{u}}\right)^{\text{T}} + [\mathcal{V}]_{(2)}, \\
[\mathcal{Q}]_{(3)} &= \mathbf{S}[\mathcal{R}]_{(3)}\left(\mathbf{G}^{\text{T}} \otimes \mathbf{H}_{\text{d}} \otimes \mathbf{H}_{\text{u}}\right)^{\text{T}} + [\mathcal{V}]_{(3)}, \\
\label{unf4}[\mathcal{Q}]_{(4)} &= \mathbf{G}^{\text{T}}[\mathcal{R}]_{(4)}\left(\mathbf{S} \otimes \mathbf{H}_{\text{d}} \otimes \mathbf{H}_{\text{u}}\right)^{\text{T}} + [\mathcal{V}]_{(4)}, 
\end{align}
where $[\mathcal{R}]_{(1)} \in \mathbb{C}^{N \times N^5}$, $[\mathcal{R}]_{(2)} \in \mathbb{C}^{N \times N^5}$, $[\mathcal{R}]_{(3)} \in \mathbb{C}^{N^2 \times N^4}$ and $[\mathcal{R}]_{(4)} \in \mathbb{C}^{N^2 \times N^4}$ are the unfoldings of the core tensor, while $[\mathcal{V}]_{(1)} \in \mathbb{C}^{M \times MKL}$, $[\mathcal{V}]_{(2)} \in \mathbb{C}^{M \times MKL}$, $[\mathcal{V}]_{(3)} \in \mathbb{C}^{K \times M^{2}L}$ and $[\mathcal{V}]_{(4)} \in \mathbb{C}^{L \times M^{2}K}$ represent the unfoldings of the noise tensor.

\section{Proposed Two-Stage CE Algorithm}
\subsection{Stage 1: Trilinear Alternating Least Square (TALS)}
At the BS, from (\ref{t4}), the final goal is to independently estimate in a joint manner the non-reciprocal channels $\{\mathbf{H}_{\text{d}}$, $\mathbf{G}_{\text{d}}$, $\mathbf{H}_{\text{u}}$, $\mathbf{G}_{\text{u}}\}$. This is achieved by solving the following multilinear optimization problem:
\begin{equation}\label{op}
\underset{\mathbf{H}_{\text{u}}, \mathbf{H}_{\text{d}}, \mathbf{G}}{\text{min}}\left\|\mathcal{Q} - \mathcal{R} \times_{1} \mathbf{H}_{\text{u}} \times_{2} \mathbf{H}_{\text{d}} \times_{3} \mathbf{S} \times_{4} \mathbf{G}^{\text{T}}\right\|_{\text{F}}^{2}.
\end{equation}

To this end, the classical alternating least squares (ALS) algorithm \cite{als} can be used for estimating each factor matrix of the Tucker-4 decomposition individually by converting the multilinear optimization problem in (\ref{op}) into three independent and simplest LS sub-problems by exploiting the unimodal unfoldings of $\mathcal{Q}$. In more detail, this solution consists in fitting iteratively a Tucker-4 decomposition to the received signal tensor from Equations (\ref{unf1}), (\ref{unf2}) and (\ref{unf4}) by minimizing the residual error (i.e., the error between the received noisy tensor $\mathcal{Q}$ and its reconstructed version $\hat{\mathcal{Q}}$ computed from the estimated factor matrices $\hat{\mathbf{H}}_{\text{u}}$, $\hat{\mathbf{H}}_{\text{d}}$ and $\hat{\mathbf{G}}$). This is done in a trilinear alternately way in which one given factor matrix is updated by fixing the other matrices to their values obtained at previous updating steps. At each iteration of the \textit{Stage 1} of the CE algorithm, the following LS problems are alternately solved
\begin{align*}
\hat{\mathbf{H}}_{\text{u}} &= \underset{\mathbf{H}_{\text{u}}}{\text{argmin}}\left\|[\mathcal{Q}]_{(1)} - \mathbf{H}_{\text{u}}[\mathcal{R}]_{(1)}\left(\mathbf{G}^{\text{T}} \otimes \mathbf{S} \otimes \mathbf{H}_{\text{d}}\right)^{\text{T}}\right\|_{\text{F}}^{2},\\
\hat{\mathbf{H}}_{\text{d}} &= \underset{\mathbf{H}_{\text{d}}}{\text{argmin}}\left\|[\mathcal{Q}]_{(2)} - \mathbf{H}_{\text{d}}[\mathcal{R}]_{(2)}\left(\mathbf{G}^{\text{T}} \otimes \mathbf{S} \otimes \mathbf{H}_{\text{u}}\right)^{\text{T}}\right\|_{\text{F}}^{2},\\
\hat{\mathbf{G}}^{\text{T}} &= \underset{\mathbf{G}}{\text{argmin}}\left\|[\mathcal{Q}]_{(4)} - \mathbf{G}^{\text{T}}[\mathcal{R}]_{(4)}\left(\mathbf{S} \otimes \mathbf{H}_{\text{d}} \otimes \mathbf{H}_{\text{u}}\right)^{\text{T}}\right\|_{\text{F}}^{2}.
\end{align*}
The solutions of these problems are given, respectively, by

\begin{align}
\hat{\mathbf{H}}_{\text{u}} &= [\mathcal{Q}]_{(1)}\left[[\mathcal{R}]_{(1)}\left(\mathbf{G}^{\text{T}} \otimes \mathbf{S} \otimes \mathbf{H}_{\text{d}}\right)^{\text{T}}\right]^{\dag},\\
\label{hd}\hat{\mathbf{H}}_{\text{d}} &= [\mathcal{Q}]_{(2)}\left[[\mathcal{R}]_{(2)}\left(
\mathbf{G}^{\text{T}} \otimes \mathbf{S} \otimes \mathbf{H}_{\text{u}}\right)^{\text{T}}\right]^{\dag},\\
\label{g}\hat{\mathbf{G}}^{\text{T}} &=[\mathcal{Q}]_{(4)}\left[[\mathcal{R}]_{(4)}\left(\mathbf{S} \otimes \mathbf{H}_{\text{d}} \otimes \mathbf{H}_{\text{u}}\right)^{\text{T}}\right]^{\dag}.
\end{align}

These three LS update steps are repeated until convergence, which is declared when $|\epsilon_{(i)} - \epsilon_{(i-1)}| \leq 10^{-6}$, where $\epsilon_{(i)} = \|\mathcal{Q} - \hat{\mathcal{Q}}_{(i)} \|_{\text{F}}^{2}$ and $\hat{\mathcal{Q}}_{(i)} = \mathcal{R} \times_{1} \hat{\mathbf{H}}_{\text{u}(i)} \times_{2} \hat{\mathbf{H}}_{\text{d}(i)} \times_{3} \mathbf{S} \times_{4} \hat{\mathbf{G}}^{\text{T}}_{(i)}$ is the reconstructed version of $\mathcal{Q}$ obtained from the estimated factor matrices at the end of the $i$-th iteration. Note that in the context of this work, $\mathcal{R}$ and $\mathbf{S}$ are known at the BS. 

\subsection{Stage 2: Khatri-Rao Factorization (KRF)}
The TALS stage provides estimates of $\hat{\mathbf{H}}_{\text{d}}$ and $\hat{\mathbf{H}}_{\text{u}}$. However, the estimates of $\hat{\mathbf{G}}_{\text{d}}$ and $\hat{\mathbf{G}}_{\text{u}}$ are not directly obtained. Instead, the structured matrix $\hat{\mathbf{G}} = \hat{\mathbf{G}}^{\text{T}}_{\text{d}} \diamond \hat{\mathbf{G}}^{\text{T}}_{\text{u}}$ is given as output of \textit{Stage 1}. In order to obtain individually $\hat{\mathbf{G}}_{\text{d}}$ and $\hat{\mathbf{G}}_{\text{u}}$ from $\hat{\mathbf{G}}$, the optimization problem $\underset{\hat{\mathbf{G}}_{\text{d}},\hat{\mathbf{G}}_{\text{u}}}{\text{min}}\left\|\hat{\mathbf{G}} - \hat{\mathbf{G}}^{\text{T}}_{\text{d}} \diamond \hat{\mathbf{G}}^{\text{T}}_{\text{u}}\right\|_{\text{F}}^{2}$ should be solved. Its state-of-the-art solution comes from the KRF method \cite{7077557} that solves in a closed-form way multiple rank-one matrix approximation problems \cite{9361077}. The steps of proposed two-stage CE algorithm are summarized in Algorithm 1. 

\begin{algorithm}[ht]
    \small
	\caption{Proposed Two-Stage CE Algorithm}
	\label{Alg1}
	\begin{algorithmic}
			\STATE \hspace{-4ex} 1. Set $i=0$ and initialize $\hat{\mathbf{G}}_{(i=0)}$ randomly;\\
			\STATE \hspace{-4ex} 2. $i \leftarrow i + 1$;\\
			\STATE \hspace{-4ex} 3. Get $\hat{\mathbf{H}}_{\text{u}(i)} = [\mathcal{Q}]_{(1)}\Bigl[[\mathcal{R}]_{(1)}\bigl(\hat{\mathbf{G}}_{(i-1)}^{\text{T}} \otimes \mathbf{S} \otimes \hat{\mathbf{H}}_{\text{d}(i-1)}\bigr)^{\text{T}}\Bigr]^{\dag}$;\\
           \STATE \hspace{-4ex} 4. Get $\hat{\mathbf{H}}_{\text{d}(i)} = [\mathcal{Q}]_{(2)}\Bigl[[\mathcal{R}]_{(2)}\bigl(\hat{\mathbf{G}}_{(i-1)}^{\text{T}} \otimes \mathbf{S} \otimes \hat{\mathbf{H}}_{\text{u}(i)}\bigr)^{\text{T}}\Bigr]^{\dag}$;\\
           \STATE \hspace{-4ex} 5. Get $\hat{\mathbf{G}}_{(i)}^{\text{T}} =[\mathcal{Q}]_{(4)}\Bigl[[\mathcal{R}]_{(4)}\bigl(\mathbf{S} \otimes \hat{\mathbf{H}}_{\text{d}(i)} \otimes \hat{\mathbf{H}}_{\text{u}(i)}\bigr)^{\text{T}}\Bigr]^{\dag}$;\\
           \STATE \hspace{-4ex} 6. Compute the residual error $\epsilon_{(i)} = \|\mathcal{Q} - \hat{\mathcal{Q}}_{(i)}\|_{\text{F}}^{2}$, where
           \STATE $\hat{\mathcal{Q}}_{(i)} = \mathcal{R} \times_{1} \hat{\mathbf{H}}_{\text{u}(i)} \times_{2} \hat{\mathbf{H}}_{\text{d}(i)} \times_{3} \mathbf{S} \times_{4} \hat{\mathbf{G}}^{\text{T}}_{(i)}$;
           \STATE \hspace{-4ex} 7. Repeat steps 2-6 until convergence;\\
           \STATE \hspace{-4ex} 8. From $\hat{\mathbf{G}}$, obtain $\hat{\mathbf{G}}_\mathrm{d}$ and $\hat{\mathbf{G}}_\mathrm{u}$ applying KRF \cite{7077557}.\\
           \STATE \hspace{-4ex} 9. Return $\hat{\mathbf{H}}_{\text{d}}$, $\hat{\mathbf{G}}_{\text{d}}$, $\hat{\mathbf{H}}_{\text{u}}$ and $\hat{\mathbf{G}}_{\text{u}}$.
	\end{algorithmic}
\end{algorithm}

\subsection{Identifiability}
Uniqueness in the LS sense of $\{\hat{\mathbf{H}}_{\text{d}}$, $\hat{\mathbf{G}}_{\text{d}}$, $\hat{\mathbf{H}}_{\text{u}}$, $\hat{\mathbf{G}}_{\text{u}}\}$ are guaranteed when $T \geq M$ related to pilot matched filtering in (\ref{signalPart}), $P \geq L$ for the coding matched filtering in (\ref{tot}), $LKM \geq N$ and $KM^2 \geq N^2$ to calculate the pseudo-inverses in steps 3, 4, and 5 in Algorithm 1. The estimated matrices are affected by simple column scaling ambiguities, which can be removed through normalization procedure \cite{hu2021two}. 

\section{Simulation Results}
\begin{figure}[t]
	\centering
	\includegraphics[width=0.43\textwidth]{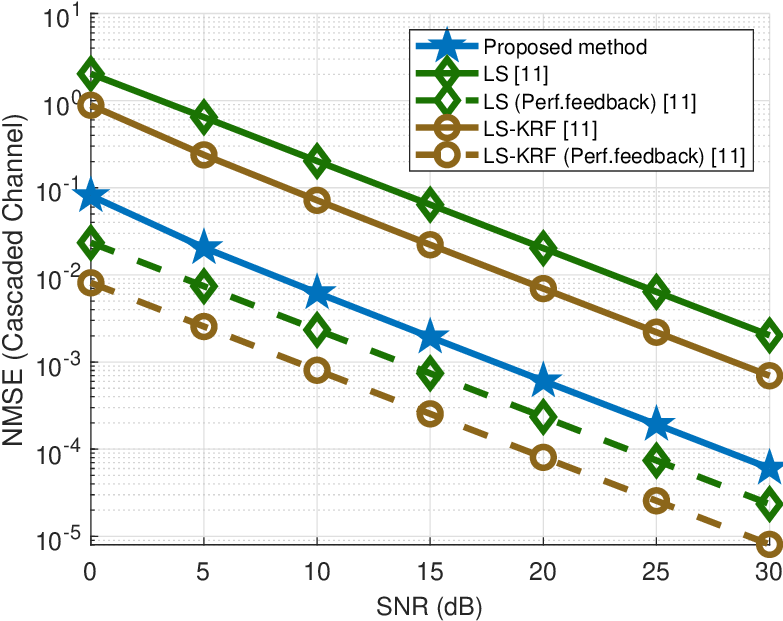}
	\caption{NMSE performance of cascaded channels versus SNR.}
	\label{fig:nmsegtheta}
\end{figure}
\begin{figure}[t]
	\centering
	\includegraphics[width=0.43\textwidth]{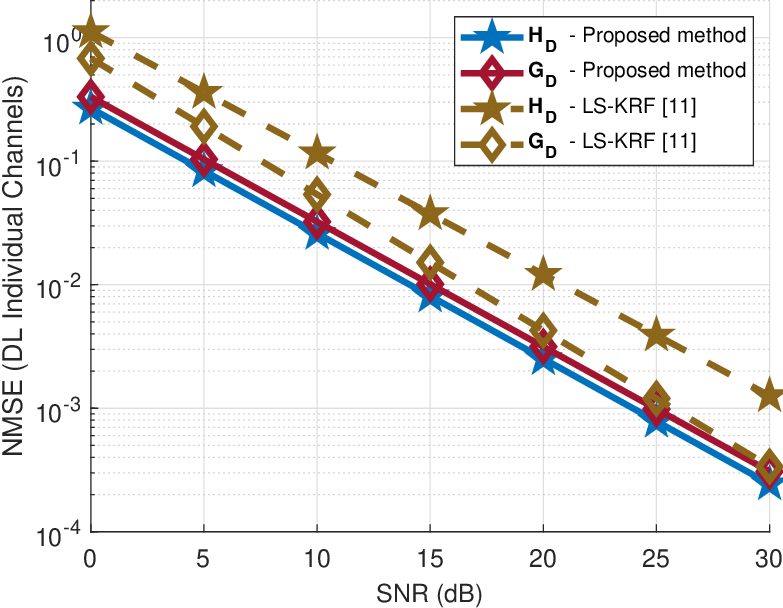}
	\caption{NMSE performance of DL individual channels versus SNR.}
	\label{fig:nmsedownlink}
\end{figure}
\begin{figure}[t]
	\centering
	\includegraphics[width=0.43\textwidth]{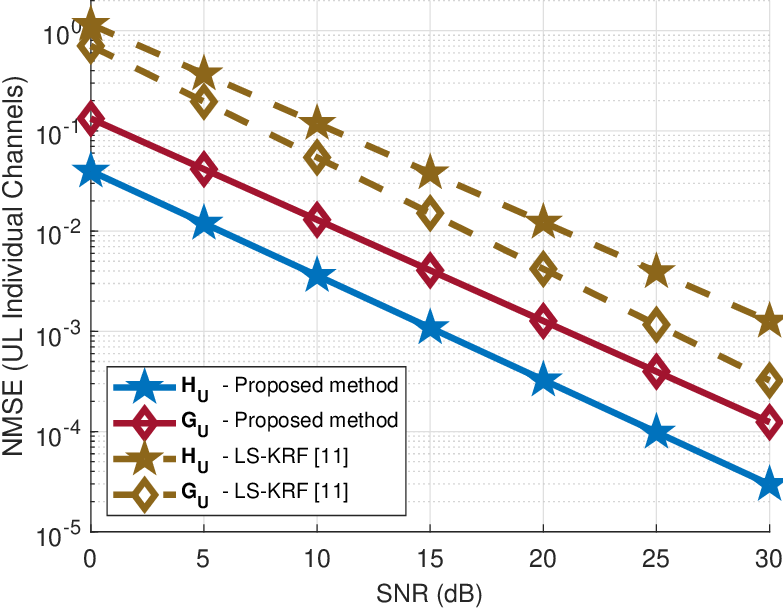}
	\caption{NMSE performance of UL individual channels versus SNR.}
	\label{fig:nmseuplink}
\end{figure}

We evaluate the performance of Algorithm 1 in terms of normalized mean squared error (NMSE), defined as $\mathrm{NMSE}(\mathbf{\Omega}) = \|\mathbf{\Omega}- \hat{\mathbf{\Omega}}\|_\mathrm{F}^2 / \|\mathbf{\Omega}\|_\mathrm{F}^2$, where $\mathbf{\Omega} \in \{\mathbf{H}_{\text{d}}$, $\mathbf{G}_{\text{d}}$, $\mathbf{H}_{\text{u}}$, $\mathbf{G}_{\text{u}}\}$. The channel matrices are Rayleigh fading, whose entries follow a zero-mean independent and identically distributed (i.i.d.) complex-valued Gaussian distribution. For simplicity, we assume that BS and UT present the same signal-to-noise ratio (SNR) level in the pilot reception. All plots were obtained by averaging $10^3$ independent Monte Carlo runs, with different channels and noise realizations for each of them. The system parameters are set to $\{M,L,N,T,K,P\}$ = $\{8,4,16,16,64,8\}$. The pilot, coding, and DL RIS response matrices are designed as truncated discrete Fourier transform matrices, while the UL RIS response is a truncated Hadamard one. The traditional matrix-based LS estimator and the tensor-based LS-KRF method proposed by \cite{9361077}, both operating in FDD mode, are considered benchmarks. In these approaches, the BS transmits pilots in the DL. Then, the UT performs CE using LS or LS-KRF estimators. The estimated DL channel is then fed back to the BS through dedicated feedback resources. Similarly, the BS estimates the UL channel and fed back to the UT. On the other hand, the proposed method concentrates completely on the joint DL and UL CE tasks at the BS side, eliminating the need for dedicated feedback channels as well as reducing the UT processing complexity. 

Fig. \ref{fig:nmsegtheta} shows the NMSE for the estimated cascaded channels. In the benchmark methods, we consider noise-free and imperfect feedback of the estimated channels. The first one represents the \textit{clairvoyant} lower bound estimator (ideal case) that does not correspond to a practical application. In the second one, the feedback channel is modeled as an AWGN channel with the same SNR considered in the DL and UL pilots' reception \cite{7353214}. From this figure and as expected, the estimation accuracy linearly varies as a function of the SNR for all simulated methods. It can be seen that the LS-KRF estimator outperforms the LS one. This result corroborates with \cite{9361077}, since the rank-one matrix approximation performed by LS-KRF provides denoising in the cascaded channel estimation compared to LS estimator which works in a single and direct estimation stage. Note also that when imperfect feedback is assumed, the proposed method achieves substantial performance improvement over both benchmark approaches while reducing CE processing at the UT side. This result illustrates the ability of the proposed tensor-based method to accurately estimate the cascaded channel even when both DL and UL noises are considered simultaneously in the received signal. Similar to the LS-KRF estimator in \cite{9361077}, the proposed method also provides some denoising level since consecutive matrix rank-one approximations are computed in \textit{Stage 2}.

Figs. \ref{fig:nmsedownlink} and \ref{fig:nmseuplink} depict the CE performance of the individual DL and UL non-reciprocal channels by assuming imperfect feedback to the LS-KRF estimator. This result illustrates the ability of the proposed method to estimate individually four non-reciprocal channels with high accuracy. We can note that the estimates of $\{\mathbf{G}_{\text{d}}, \mathbf{G}_{\text{u}}\}$ are worse compared to $\{\mathbf{H}_{\text{d}}, \mathbf{H}_{\text{u}}\}$. This is due to the error propagation since the individual estimates of the DL and UL RIS-UT channels are obtained at \textit{Stage 2} of the proposed method that suffers from estimation errors coming from \textit{Stage 1}. The individual estimates of the non-reciprocal channels are important in particular applications, for instance, user localization, channel sounding, mobility tracking, and scenarios in which the DL and UL RIS-UT channels vary faster than the DL and UL BS-RIS. 

\section{Conclusion}
We presented a novel joint DL-UL CE approach for non-reciprocal RIS-assisted communications. Through a closed-loop three-phase protocol, we capitalized on the Tucker tensor decomposition to model the received signal, from which all the involved channel matrices can be individually estimated using a sequential combination of iterative and closed-form estimation procedures. The proposed two-stage TALS/KRF-based algorithm is designed to concentrate the processing burden for CE at the BS while relaxing non-reciprocity assumptions involving the DL and UL RIS-assisted channels. \textcolor{black}{Our simulation results have illustrated the high CE accuracy of the proposed joint UL-DL channel estimation algorithm.} As perspectives for future work, we will tackle the joint passive/active beamforming optimization problem for non-reciprocal RIS based on our tensor modeling formulation.  

\bibliographystyle{IEEEtran}
\bibliography{ref}

\end{document}